 \documentclass[preprint,review,12pt]{elsarticle}

\usepackage{amssymb}

\journal{Communications in Nonlinear Science and Numerical Simulation}

\begin{document}

\begin{frontmatter}



\title{Thermodynamics of a bouncer model: a simplified one-dimensional gas}


\author{Edson D.\ Leonel$^{1,2}$}
\ead{edleonel@rc.unesp.br}
\author{Andr\'e L. P. Livorati$^{3,4}$}
\address{$^1$UNESP - Univ Estadual Paulista - Departamento de F\'isica,
Av.24A 1515 - Bela Vista - 13506-900 - Rio Claro - SP - Brazil - Tel.:
+55-19-3526 9174 - Fax: +55-19 3526 9181\\
$^2$ The Abdus Salam - ICTP, Strada Costiera, 11 - 34151 - Trieste -
Italy\\
$^3$ Instituto de F\'isica, Univ S\~ao Paulo, Rua do Mat\~ao, 
Cidade Universit\'aria - 05314-970 - S\~ao Paulo - SP - Brazil\\ 
$^4$ School of Mathematics, University of Bristol, Bristol BS8 1TW, United
Kingdom}

\begin{abstract}
Some dynamical properties of non interacting particles in a bouncer model are
described. They move under gravity experiencing collisions with a moving
platform. The evolution to steady state is described in two cases for
dissipative dynamics with inelastic collisions: (i) for large initial energy;
(ii) for low initial energy. For (i) we prove an exponential decay while for
(ii) a power law marked by a changeover to the steady state is observed. A
relation for collisions and time is obtained and allows us to write relevant
observables as temperature and entropy as function of either number of
collisions and time.
\end{abstract}

\begin{keyword}
Bouncer model \sep Diffusion in energy \sep Scaling
\end{keyword}

\end{frontmatter}



\section{Introduction}
\label{intro}

~~~Modelling a dynamical system has become one of the most challenging
subjects
among scientists including physicists and mathematicians over years
\cite{Ref1,Ref2}. The modelling helps to understand in many cases how does
the system evolves in time \cite{Ref3} was well as its description in
parameter space \cite{Ref4,Ref5} and whether it has or not a steady state
\cite{Ref6}. Often the investigation leads to nonlinear dynamics \cite{Ref7}
where complex structures can be observed in the phase space \cite{Ref8}. For
conservative systems, the phase space may be classified under three different
classes namely: (i) regular \cite{Ref9} where only periodic and quasi periodic
orbits are present; (ii) mixed \cite{Ref10} whose phase space exhibits a
coexistence of period, quasi periodic and chaotic behaviour and; (iii) ergodic
\cite{Ref11} where only unstable and therefore unpredictable orbits
are observed. For dissipative systems \cite{Ref12} the structure of the phase
space commonly has attractors \cite{Ref13} that can be periodic \cite{Ref14}
or chaotic \cite{Ref15}.

In the large majority of the cases, a dynamical system is mostly governed by a
set of differential equations. Quite often too they are coupled to each other.
However, depending on the conserved quantities and symmetries, solutions of
differential equations can be qualitatively (and many times quantitatively
too) transformed into an application described by nonlinear mappings
\cite{Ref16}. The mappings are characterised by discrete time evolution and
have also a set of control parameters. Indeed they can control either the
nonlinearity \cite{Ref17} as well as the dissipation itself \cite{Ref18}.

The variation of the control parameters may lead quite often to the so called
phase transitions \cite{Ref19,Ref20}. In statistical mechanics, phase
transitions are linked to abrupt changes in spatial structure of the system
\cite{Ref21,Ref22} and mainly due to variations of control parameters. In a
dynamical system however, a phase transition is particularly related to
modifications in the structure of the phase space of the system
\cite{Ref23,Ref24}. Therefore near a phase transition, the dynamics of the
system is described by the use of a scaling function \cite{Ref25,Ref26} where
critical exponents characterise the dynamics near the criticality.

In this paper we revisit a bouncer model \cite{Ref27,Ref28} particularly
focused on the description of some of its thermodynamical properties. The
system consists of a classical particle, or in the same way an ensemble of
non interacting particles, moving under the action of a constant gravitational
field and suffering collisions with a moving platform. We are seeking then to
understand and describe how does the system goes to the steady state for long
enough time and how the control parameters influence the way the system goes.
For the conservative case and depending on the control parameters
\cite{Ref29}, the system exhibits unlimited diffusion in energy \cite{Ref30}
which is called as Fermi acceleration \cite{Ref31}. If the particles are
considered as a sufficiently light, an ensemble of them may constitute an
ideal gas. Therefore the unlimited diffusion in energy is in contrast with
what is observed in day life. If we consider the moving wall as produced by
atomic oscillations in a solid due to thermal heating with a constant external
temperature $T_e$, a gas in a room does not absorbs infinite energy leading it
to have unbound growth of temperature. Therefore Fermi acceleration is a
phenomenon that can not occur in gases and most probably is due to the fact
that dissipation is present. Because the gas is of low density and particles
are non interacting with each other, we consider the dissipation is due to
inelastic collisions of the particles with the moving wall, but the particles
do not interact between themselves. When interactions among the particles are
considered, the system can be described as a granular material \cite{add1}
allowing physical observables to be characterised \cite{add2,add3} either in
the presence \cite{add4} or absence \cite{add5} of gravitational field. Such
approach is not considered in this paper given we are considering non
interaction particles with low density. In model considered in this paper we
introduce a dissipation parameter, more specifically a restitution
coefficient, and describe the evolution of the system. Then the introduction
of inelastic collisions of the particle with the wall suppresses the diffusion
in energy. The evolution towards the stationary state for long enough time is
described in two limits: (i) if the initial energy of the gas is sufficiently
large and; (ii) if it is sufficiently small. For case (i) we prove an
exponential decay is happening while for (ii) a power law marked by a
changeover to the steady state is observed. We obtain so far a relation of the
number of collisions and time and write the relevant observables like squared
velocity, temperature of the gas and entropy as a function of either number of
collisions and time. The system is show to be scaling invariant with respect
to the control parameters and we found analytically the critical exponents
describing a homogeneous generalised function. At the end we present some
comparisons of the results obtained with typical values of known atomic
oscillations as well as frequency of oscillation at a given temperature. Our
results indicate the inelastic collisions are responsible for suppressing the
energy's unlimited diffusion of the gas. An estimation of a restitution
coefficient is given for Hydrogen molecule colliding with a solid made of
copper.

This paper is organised as follows. In Sec. \ref{sec2} we describe the
model, give the expressions of the conservative and dissipative maps.
Unlimited diffusion for energy is shown also for the conservative case. The
stationary state and the evolution towards it is given here as a theoretical
prediction. Section \ref{sec3} is devoted to discuss the numerical results as
well as the scaling properties. The critical exponents are obtained from
either theoretical point of view as well as from numerical simulations. A
scaling invariance for the ensemble of particles in a gas is confirmed by an
overlap of different curves of average velocity onto a single and universal
plot, after properly rescaling of the axis. As the system is constructed and
described in terms of the number of collisions, section \ref{sec4} deals
specifically with a connection of number of collisions and time. The latter
being in principle easier to be measured in a real experiment. Section
\ref{sec5} discusses the connections with the thermodynamics finding
particularly the expressions of the temperature of the gas, squared velocity
as well as an expression for the entropy. Both as function of number of
collisions as well as the time. Short discussion on the results are presented
in section \ref{sec6} where, from our model, an estimation of the restitution
coefficient for collisions of Hydrogen molecule is given. Conclusions are
presented in section \ref{sec7}.

\section{The model and the map}
\label{sec2}

\begin{figure}[htb]
\centerline{\includegraphics[width=0.55\linewidth]{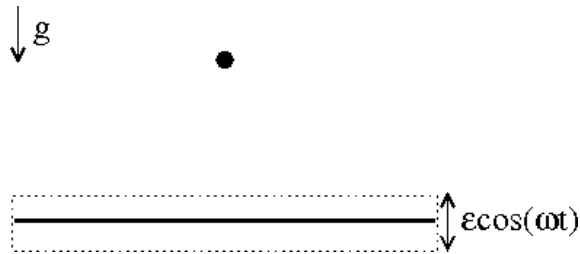}}
\caption{{\it Schematic description of the model.}}
\label{Fig1}
\end{figure}

~~~The model we consider in this paper consists of a classical particle, or an
ensemble of non interacting particles, moving in the presence of a constant
gravitational field $g$ suffering collisions with a time moving wall. The
equation that describes the moving wall is given by
\begin{equation}
y_w(t)=\epsilon\cos(\omega t)~,
\label{Eq1}
\end{equation}
where $\epsilon$ denotes the amplitude of the moving wall while $\omega$ is
the angular frequency. We suppose the gas of particles has a low density in
the sense the particles are free to move with a constant mechanical energy
without interacting with each other. They indeed exchange energy upon
collisions with the moving wall. Depending on the phase of the moving wall the
particles can gain or lose energy. It is assumed the motion of the particles
is allowed only in the vertical direction, therefore making allusions to a
simplified one-dimensional gas. Figure \ref{Fig1} shows a schematic
description of the model.

The dynamics of each individual particle, as usual in the literature
\cite{Ref2}, is described by a two-dimensional and nonlinear map for the
variables velocity of the particle $V$ and time $t$ at each impact with the
boundary. For a simplified gas model, the moving platform may be represented
by a rigid wall or even the ground and the motion can be caused by the atomic
oscillations at the edge of the wall. Because such atomic oscillations are too
small \cite{Ref32} as compared to the positions or displacements made
by each particle, we can approximate the description by assuming the time of
flight of each particle is calculated as if the wall was fixed. However, the
exchange of energy is determined by a moving wall. This approach is
reasonable because the atomic oscillations are from the order of $10^{-11}m$
(see \cite{Ref32}). Under this assumption, the mapping is written as
\begin{equation}
\left\{\begin{array}{ll}
t_{n+1}=\left[t_n+2{{V_n}\over{g}}\right]~{\rm mod~(2 \pi/\omega)} \\ 
V_{n+1}=|\gamma V_n-(1+\gamma)\epsilon\omega\sin(\omega t_{n+1})|
\end{array}
\right..
\label{Eq2}
\end{equation}
The modulus used in the second equation is introduced to avoid a specific
situation. After the collisions, there could be a small possibility of the
particle present a negative velocity. This case has no physical meaning in
the model because the wall is assumed to be fixed in order to make easier the
calculation of the time of flights. Then a particle moving with a negative
velocity after the collision is forbidden. If such a case happens, the
particle is injected back to the dynamics with the same velocity but with a
positive direction. The parameter $\gamma$ denotes the restitution coefficient
for the impacts. Indeed $\gamma\in[0,1]$. As we will see, for $\gamma=1$ the
conservative dynamics is observed. For a specific range of control parameters,
this leads to an unlimited diffusion in velocity \cite{Ref29} therefore
characterising a phenomenon called as Fermi acceleration \cite{Ref31}. As a
physical interpretation from a thermodynamical point of view, it would leads
to an infinite temperature of a classical gas. This is quite contradictory to
what is observed in real experiments or in day life. On the other hand when
inelastic collisions are taken into account, the unlimited diffusion of the
velocity of the ensemble of particles is suppressed, leading to a finite
temperature most in agreement to what is confirmed in experiments. We discuss
both cases separately.

\subsection{Conservative dynamics}
\label{sec2.1}

~~~Let us discuss in this section the behaviour of the average velocity for an
ensemble of particles for the restitution coefficient $\gamma=1$. Under this
condition, the mapping simplifies to
\begin{equation}
\left\{\begin{array}{ll}
t_{n+1}=\left[t_n+2{{V_n}\over{g}}\right]~{\rm mod~(2 \pi/\omega)} \\ 
V_{n+1}=|V_n-2\epsilon\omega\sin(\omega t_{n+1})|
\end{array}
\right..
\label{Eq3}
\end{equation}
The expression for mapping (\ref{Eq3}) is remarkably similar to the one
describing the standard map \cite{Ref2} (please see appendix 1). There are few
steps we have to proceed to make a connection between the two models: (1)
multiply first equation of map (\ref{Eq3}) by $\omega$; (2) multiply the
second equation of map (\ref{Eq3}) by $2\omega/g$; (3) define both
$\tilde{\phi}=\omega t$ and $I=2\omega V/g$; (4) define
$\phi_{n+1}=\tilde{\phi}_n+\pi$. After these steps, we obtain an effective
control parameter as
\begin{equation}
K_{\rm eff}={{4\epsilon\omega}\over{g}}~.
\label{Eq4}
\end{equation}
Therefore for $K_{\rm eff}>K_c$ the phase space allows to unlimited diffusion
in the velocity \cite{Ref29}. To make sure we are considering such case, in
our numerical simulations we shall consider only
\begin{equation}
{{\epsilon\omega}\over{g}}\ge 0.2429\ldots~.
\label{Eq5}
\end{equation}

To give an analytical argument on the unlimited diffusion, let us start the
ensemble of particles with a low initial velocity. We mean low here an
ensemble with low temperature but large enough so that quantum effects can be
disregarded. Then squaring second equation of mapping (\ref{Eq3}) we obtain
\begin{equation}
V_{n+1}^2=V_n^2-4\epsilon\omega V_n\sin(\omega
t_{n+1})+4\epsilon^2\omega^2\sin^2(\omega t_{n+1})~.
\label{Eq6}
\end{equation}

\begin{figure}[htb]
\centerline{\includegraphics[width=0.55\linewidth]{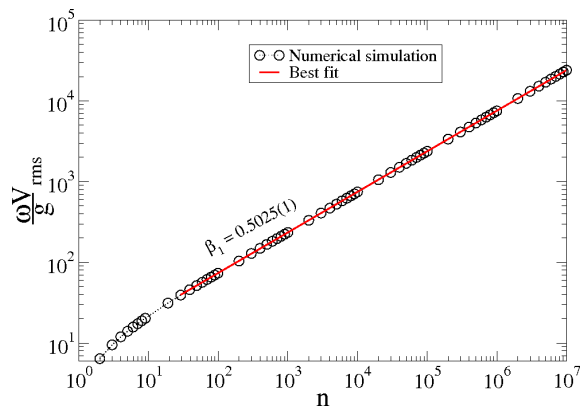}}
\caption{{\it Plot of the average velocity as a function of $n$. A power law
fitting furnishes a slope of $0.5025(1)\cong1/2$, in remarkably well
agreement with Eq. (\ref{Eq8}). The parameters used were such that
$(\epsilon\omega^2/g)=10$.}}
\label{Fig2}
\end{figure}

Taking an average over an ensemble of different initial times
$t\in[0,2\pi/\omega]$, we end up with
\begin{eqnarray}
\overline{V^2}_{n+1}-\overline{V^2}_n&=&{{\overline{V^2}_{n+1}-\overline{V^2}
_n}\over{(n+1)-n}}~, \nonumber\\
&\cong&{{\partial \overline{V^2}}\over{\partial n}}=2\epsilon^2\omega^2~.
\label{Eq7}
\end{eqnarray}
We have then a differential equation involving $\overline{V^2}$ and $n$.
Doing the integration properly we obtain that
\begin{eqnarray}
V_{\rm rms}&=&\sqrt{\overline{V^2}}~,\nonumber\\
&=&\sqrt{V_0^2+2\epsilon^2\omega^2n}~.
\label{Eq8}
\end{eqnarray}
For a low initial velocity $V_0\rightarrow 0$ then we see
$V_{\rm rms}\propto n^{{1}\over{2}}$. Figure \ref{Fig2} shows a plot of the
numerical simulation made for an ensemble of $5,000$ different initial
conditions at the same initial velocity. The slope of growth is $1/2$ as
theoretically given by Eq. (\ref{Eq8}).

The result presented in Fig. \ref{Fig2} confirms the unlimited growth of the
average velocity, leading to the phenomenon of Fermi acceleration. However, it
is in disagreement with experimental results because a temperature in a room
can not grow unbound. In next section we consider the dynamics under the
effects of inelastic collisions.

\subsection{Dissipative dynamics}
\label{sec2.2}

~~~Let us discuss here the implications of the inelastic collisions in the
steady state dynamics. According to the theory of dynamical systems
\cite{Ref6}, the presence of dissipation in the system may lead to the
existence of attractors in the system. The determinant of the Jacobian matrix
of mapping (\ref{Eq1}) is $\gamma~\rm{sign}[\gamma
V_n-(1+\gamma)\epsilon\omega\sin(\omega t_{n+1})]$ where $\rm {sign} (u)=1$
for $u>0$ and $\rm {sign} (u)=-1$ for $u<0$. This is indeed a quite strong
mathematical result. According to Liouville's theorem \cite{Ref33}, area
contraction in the phase space is happening, therefore for any $\gamma<1$
attractors must exist in the phase space. Given they are far away from the
infinity, unlimited diffusion must not be observed anymore. Our results
corroborate with this. Moreover if the dissipation suppresses the unlimited
diffusion in velocity, Fermi acceleration is suppressed too making us to
reinforce that it is not a robust phenomena \cite{Ref34}. 

To make some theoretical progress, we start squaring the second equation of
mapping (\ref{Eq2}), that leads to
\begin{equation}
V^2_{n+1}=\gamma^2V^2_n-2\gamma V_n(1+\gamma)\epsilon\omega\sin(\omega
t_{n+1})+(1+\gamma)^2\epsilon^2\omega^2\sin^2(\omega t_{n+1})~.
\label{Eq9}
\end{equation}
Doing an ensemble average for $t\in[0,2\pi/\omega]$ and grouping
properly the terms we obtain
\begin{equation}
\overline{V^2}_{n+1}=\gamma^2\overline{V^2}_n+{{(1+\gamma)^2\epsilon^2\omega^2
}\over{2}}~.
\label{Eq10}
\end{equation}

Equation (\ref{Eq10}) can be used in different forms. We start with
considering the stationary state.

\subsubsection{Stationary state}

~~~For the stationary state, we have that
$\overline{V^2}_{n+1}=\overline{V^2}_n=\overline{V^2}$. Substituting this
result in Eq. (\ref{Eq10}) we end up with an expression of the type
\begin{equation}
\overline{V^2}={{(1+\gamma)}\over{2}}{{\epsilon^2\omega^2}\over{(1-\gamma)}}~,
\label{Eq11}
\end{equation}
and that when applying square root from both sides
\begin{equation}
V_{\rm
rms}={\sqrt{{(1+\gamma)}\over{{2}}}}{{\epsilon\omega}\over{(1-\gamma)^{^1/_2}}
}~.
\label{Eq12}
\end{equation}
Notice that the steady state velocity does indeed depends on two terms: (i)
from the product of $\epsilon\omega$, which corresponds to the maximum
velocity of the moving wall, i.e., maximum atomic speed in the wall and; (ii)
on the inverse of the square root of the dissipation, namely
$(1-\gamma)^{-1/2}$.

\subsubsection{Evolution to the stationary state}

~~~Let us discuss here how does the average velocity of the ensemble goes to
the
equilibrium. If the initial velocity given for the ensemble is lower than the
one expressed by Eq. (\ref{Eq12}), then we should observe a regime of growth
until reaching the stationary state. One the other hand if is large enough,
there must be observed a decrease on the velocity until stationary state is
reached. As we will see, the two regimes approach the equilibrium in
different ways. We shall show the growth of the average velocity is given by a
power law while the decay of energy is remarkably fast given by an
exponential
function.

To have a glance on this we consider equation (\ref{Eq11}) rewritten in a
convenient way as
\begin{eqnarray}
\overline{V^2}_{n+1}-\overline{V^2}_n&=\Delta
V^2&={{\overline{V^2}_{n+1}-\overline{V^2}_n}\over{(n+1)-n}}\cong
{{\partial\overline{V^2}}\over{\partial n}}~,\nonumber\\
&=&\overline{V^2}(\gamma^2-1)+{{(1+\gamma)^2\epsilon^2\omega^2}\over{2}}~.
\label{Eq13}
\end{eqnarray}

Proceeding with the integration and after grouping the terms properly we
obtain
\begin{equation}
\overline{V^2}(n)=V_0^2e^{(\gamma^2-1)n}+{{(1+\gamma)\epsilon^2\omega^2}\over{
2(1-\gamma)}}\left[1-e^{(\gamma^2-1)n}\right]~.
\label{Eq14}
\end{equation}
We see from Eq. (\ref{Eq14}), and as expected, an explicit dependence on the
initial velocity, which allows us to make to distinct investigations: (1)
consider the case of $V_0\gg\epsilon\omega$ and; (ii) the case of
$V_0\ll\epsilon\omega$. We start with (i) first.

\begin{figure}[htb]
\centerline{\includegraphics[width=0.55\linewidth]{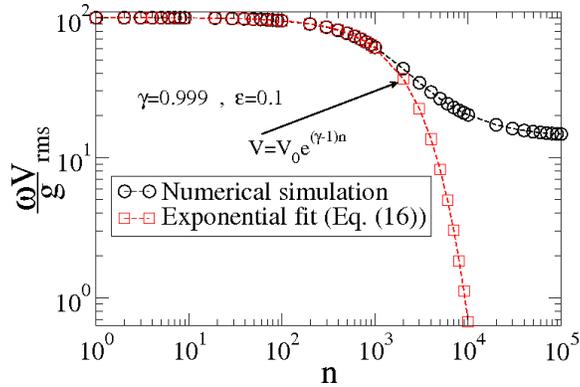}}
\caption{{\it Decay of the average velocity as function of $n$. Circles
denote the numerical simulation while squares correspond to the exponential
fit given by Eq. (\ref{Eq17}).}}
\label{Fig3}
\end{figure}

For the case of $V_0\gg \epsilon\omega$, the first term of Eq. (\ref{Eq14})
dominates over the second one. Taking square root of both sides leads to
\begin{equation}
V_{\rm rms}(n)=V_0e^{{(\gamma^2-1)n}\over{2}}~.
\label{Eq15}
\end{equation}
The numerator heading the exponential can be factored and considering the
case of low dissipation, i.e. $\gamma\cong 1$ which gives $(1+\gamma)\cong
2$, therefore we obtain that
\begin{equation}
V_{\rm rms}(n)=V_0e^{(\gamma-1)n}~,
\label{Eq17}
\end{equation}
which is an exponential function as we mention before. Figure \ref{Fig3} 
shows the behaviour of the decay of the average velocity for the control
parameter $\gamma=0.999$ and $(\epsilon\omega^2/g)=0.1$, which was chosen to
make the second term slower and being possible to observe the decay
in a convenient way. We see the two curves, one marked by circles, produced by
numerical simulation, agree with the theoretical result given by Eq.
(\ref{Eq17}) and plotted as squares, for short $n$. As soon as $n$ grows and
the second term in Eq. (\ref{Eq14}) becomes considerable, the curves split
from each other.

The exponential decay can be determined also from a different way. Indeed it
emerges naturally from iteration of the mapping. From the second equation of
mapping (\ref{Eq2}) we obtain $V_1=\gamma V_0$. If we iterate
again we have $V_2=\gamma V_1=\gamma^2V_0$. For $V_3=\gamma V_2=\gamma^3
V_0$. Finally after $n$ iterations we obtain
\begin{equation}
V_n=\gamma^nV_0~.
\label{Eq18}
\end{equation}
We can then expand Eq. (\ref{Eq18}) in Taylor series and obtain
\begin{eqnarray}
V_n&=&V_0\left[1+(\gamma-1)n+{{1}\over{2}}(\gamma-1)^2n(n-1)+{{1}\over{6}}
(\gamma-1)^3n(n-1)(n-2)\right]+\nonumber\\
&+&V_0\left[{{1}\over{24}}(\gamma-1)^4n(n-1)(n-2)(n-3)+\ldots \right]~.
\label{Eq19}
\end{eqnarray}
For $n\gg 1$, we can rewrite Eq. (\ref{Eq19}) as
\begin{equation}
V_n=V_0\left[1+(\gamma-1)n+{{1}\over{2}}(\gamma-1)^2n^2+{{1}\over{6}}
(\gamma-1)^3n^3+{{1}\over{24}}(\gamma-1)^4n^4+\ldots \right]~,
\label{Eq20}
\end{equation}
which is the own definition of the exponential given by Eq. (\ref{Eq17}).

\begin{figure}[htb]
\centerline{\includegraphics[width=0.55\linewidth]{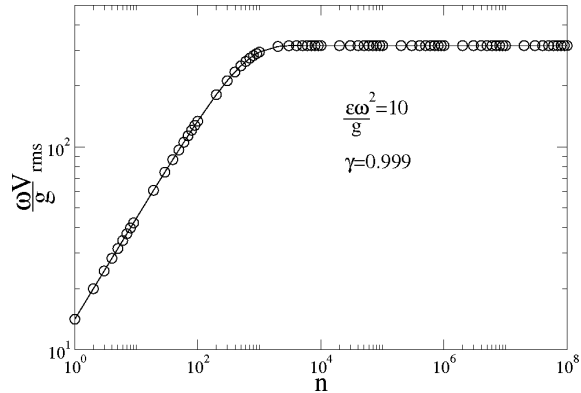}}
\caption{{\it Plot of the average velocity as function of $n$ for the case of
$V_0\ll\epsilon\omega$.}}
\label{Fig4}
\end{figure}

For the case of $V_0\ll\epsilon\omega$, the average velocity grows and
eventually bends towards a regime of saturation, which is marked by a
constant plateau for long enough $n$, as can be seen from Fig. \ref{Fig4}.
The curve shown in Fig. \ref{Fig4} grows with $n^{1/2}$ and, after reaching a
typical number of collisions marking a crossover, it bends in direction of
the stationary state. The behaviour shown in Fig. \ref{Fig4} has many more
important properties which we address in the next section, in particular a
scaling behaviour.

\section{Numerical results and scaling properties}
\label{sec3}

~~~In this section we discuss with more details the behaviour observed in Fig.
\ref{Fig4}. It is however more convenient to do numerical simulations if we
define a set of dimensionless variables. We then define
\begin{equation}
v={{\omega V}\over{g}}~~~,~~~\phi=\omega
t~~~,~~~\varepsilon={{\epsilon\omega^2}\over{g}}~,
\label{Eq21}
\end{equation}
where the parameter $\varepsilon$ corresponds to the ratio of the maximum
acceleration of the wall by the acceleration of the gravity. In this new set
of variables the map is written as
\begin{equation}
\left\{\begin{array}{ll}
\phi_{n+1}=[\phi_n+2v_n]~{\rm mod~(2 \pi)} \\ 
v_{n+1}=|\gamma v_n-(1+\gamma)\varepsilon\sin(\phi_{n+1})|
\end{array}
\right..
\label{Eq22}
\end{equation}

Let us give a short discussion on the most relevant control parameter of the
three originally involved. Indeed we assume the gravitational field $g$ is
constant. Because of the constraints of the wall as a solid body, the
amplitude of oscillation can not change much. Therefore the most significant
control parameter to the dynamics is the angular frequency $\omega$. Moreover
it appears as a power of $2$ in the definition of $\varepsilon$. Since the
atomic oscillation is really high \cite{Ref32a}, say from the order of
$10^{13}Hz$, a small modification on $\omega$ causes large variation on
$\varepsilon$.

\begin{figure}[htb]
\centerline{\includegraphics[width=0.55\linewidth]{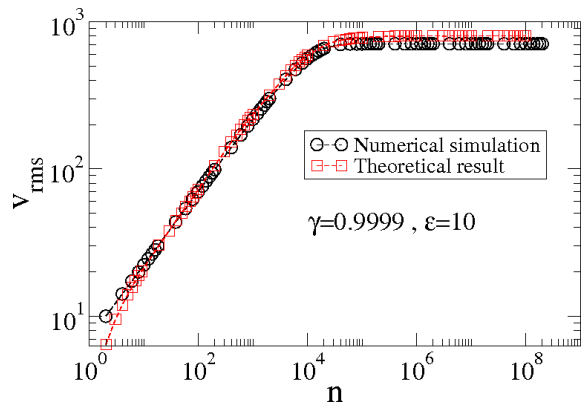}}
\caption{{\it Plot of the dimensionless average velocity as function of $n$
for the case of $V_0\ll\epsilon\omega$ and considering: simulation
(circles) and theoretical given by Eq. (\ref{Eq14}) (square).}}
\label{Fig5}
\end{figure}

Before discuss the behaviour of the average velocity as a function of the
control parameters, let us check if the theoretical result obtained from Eq.
(\ref{Eq14}) is in well agreement with the numerical simulations. It is shown
in Fig. \ref{Fig5} a plot of the the dimensionless average velocity as a
function of $n$ for both the simulation (circles) and theoretical
(squares).
We can see from the two curves a remarkable agreement, therefore letting us
to proceed with the theoretical investigation.

\begin{figure}[htb]
\centerline{\includegraphics[width=0.55\linewidth]{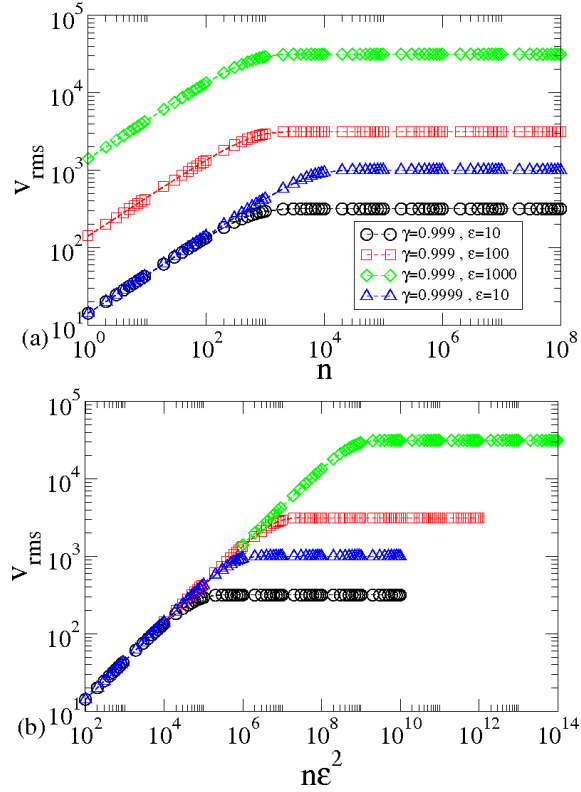}}
\caption{{\it (a) Plot of the dimensionless average velocity as function of
$n$ for the case of $V_0\ll\epsilon\omega$ and considering different control
parameters, as labelled in the figures. (b) Same plot of (a) after a
transformation $n\rightarrow n\varepsilon^2$.}}
\label{Fig6}
\end{figure}

The behaviour of the average velocity plotted against $n$ for different
control parameters is shown in Fig. \ref{Fig6}(a).
We see the curves of the average velocity start to grow with for short $n$
and, after reaching a crossover collision number $n_x$, they bend towards a
regime of saturation. Different control parameters lead the curves to
saturate at different values. The crossover $n_x$ does not seem to depend on
the parameter $\varepsilon$, as we shall confirm latter. However it does
indeed depends on the parameter $\gamma$. It is convenient to apply a
transformation $n\rightarrow n\varepsilon^2$ that makes all curves start
grow together as can be seen in Fig. \ref{Fig6}(b). This transformation
also makes the horizontal axis to have the same dimension as given by Eq.
(\ref{Eq8}).

The behaviour shown in Fig. \ref{Fig6} makes us to propose the following
scaling hypotheses:
\begin{enumerate}
\item{
\begin{equation}
\overline{v}\propto (n\varepsilon^2)^{\beta}~,{\rm~for~n\ll n_x}~,
\label{Eq23}
\end{equation}
where $\beta$ is called as the acceleration exponent;
}
\item{
\begin{equation}
\overline{v}_{\rm
sat}\propto\varepsilon^{\alpha_1}(1-\gamma)^{\alpha_2}~,~{\rm~for~n\gg n_x}~,
\label{Eq24}
\end{equation}
where $\alpha_1$ and $\alpha_2$ are the saturation exponents;
}
\item{
\begin{equation}
n_x\propto \varepsilon^{z_1}(1-\gamma)^{z_2}~,
\label{Eq25}
\end{equation}
where $z_1$ and $z_2$ are the crossover exponents.
}
\end{enumerate}
The five exponents $\beta$, $\alpha_1$, $\alpha_2$, $z_1$ and $z_2$ are
called as critical exponents.

The three scaling hypotheses shown in Eqs. (\ref{Eq23}), (\ref{Eq24}) and
(\ref{Eq25}) allow us to describe the behaviour of the average velocity in
terms of a homogeneous generalised function of the type
\begin{equation}
\overline{v}(n\varepsilon^2,\varepsilon,(1-\gamma))=\ell\overline{v}
(\ell^an\varepsilon^2,\ell^b\varepsilon,\ell^c(1-\gamma))~,
\label{Eq26}
\end{equation}
where $\ell$ is a scaling factor, $a$, $b$ and $c$ are characteristic
exponents and are related to the critical exponents.

Choosing $\ell^an\varepsilon^2=1$ we obtain
\begin{equation}
\ell=(n\varepsilon^2)^{^{-1}/_a}~.
\label{Eq27}
\end{equation}
Substituting this result in Eq. (\ref{Eq26}) we have
\begin{equation}
\overline{v}=(n\varepsilon^2)^{^{-1}/_a}\overline{v_1}((n\varepsilon^2)^{^{
-b}/_a}\varepsilon,(n\varepsilon^2)^{^{-c}/_a}(1-\gamma))~,
\label{Eq28}
\end{equation}
where the function $\overline{v_1}$ is assumed to be constant for $n\ll
n_x$. Comparing the result given by Eq. (\ref{Eq28}) with the first scaling
hypotheses given by Eq. (\ref{Eq23}), we obtain that $\beta=-1/a$. The
exponent $\beta$ was indeed confirmed as $\beta=1/2$, then $a=-2$.

Considering now $\ell^b\varepsilon=1$ we have
\begin{equation}
\ell=\varepsilon^{^{-1}/_b}~.
\label{Eq29}
\end{equation}
Substituting again this result in Eq. (\ref{Eq26}) we find
\begin{equation}
\overline{v}=\varepsilon^{^{-1}/_b}\overline{v_2}
(\varepsilon^{^{-a}/_b}(n\varepsilon^2),\varepsilon^{^{-c}/_b}
(1-\gamma))~,
\label{Eq30}
\end{equation}
where the function $\overline{v_2}$ is again assumed to be constant for $n\gg
n_x$. Comparing Eq. (\ref{Eq30}) with the second scaling hypotheses given by
Eq. (\ref{Eq24}) we obtain that $\alpha_1=-1/b$. We have already a theoretical
result for the exponent $\alpha_1$, which must be checked with numerical
simulation latter, as we will do. But a comparison with Eq. (\ref{Eq12})
yields that $\alpha_1=1$. Then scaling exponent $b=-1$.

Considering now $\ell^c(1-\gamma)=1$, we have
\begin{equation}
\ell=(1-\gamma)^{^{-1}/_c}~.
\label{Eq31}
\end{equation}
Substituting again the result obtained from Eq. (\ref{Eq31}) in Eq.
(\ref{Eq26}) we obtain
\begin{equation}
\overline{v}=(1-\gamma)^{^{-1}/_c}\overline{v_3}
((1-\gamma)^{^{-a}/_c}(n\varepsilon^2),(1-\gamma)^{^{-b}/_c}\varepsilon)~,
\label{Eq32}
\end{equation}
where the function $\overline{v_3}$ is assumed constant for $n\gg n_x$.
Comparing this result with second scaling hypotheses given by Eq.
(\ref{Eq24}), we obtain $\alpha_2=-1/c$. Theoretical result for critical
exponent $\alpha_2$ also comes from Eq. (\ref{Eq12}) and yields
$\alpha_2=-1/2$. Therefore the scaling exponent $c=2$.

The next step is to relate the critical exponents among themselves and obtain
the scaling laws. To do that, we compare the different expressions obtained
for $\ell$. First comparison is made between Eqs. (\ref{Eq27}) and
(\ref{Eq29}) and that leads to
\begin{equation}
n=\varepsilon^{{{\alpha_1}\over{\beta}}-2}~.
\label{Eq33}
\end{equation}
Equation above can be compared with third scaling hypotheses given by Eq.
(\ref{Eq25}), and that gives us
\begin{equation}
z_1={{\alpha_1}\over{\beta}}-2~.
\label{Eq34}
\end{equation}

Comparing now the expressions for $\ell$ given by Eqs. (\ref{Eq27}) and
(\ref{Eq31}) and assuming that $\varepsilon$ is constant we obtain
\begin{equation}
n=(1-\gamma)^{{\alpha_2}\over{\beta}}~,
\label{Eq35}
\end{equation}
and hence comparing with the third scaling hypotheses given by Eq.
(\ref{Eq25}) gives
\begin{equation}
z_2={{\alpha_2}\over{\beta}}~.
\label{Eq36}
\end{equation}

\begin{figure}[htb]
\centerline{\includegraphics[width=0.55\linewidth]{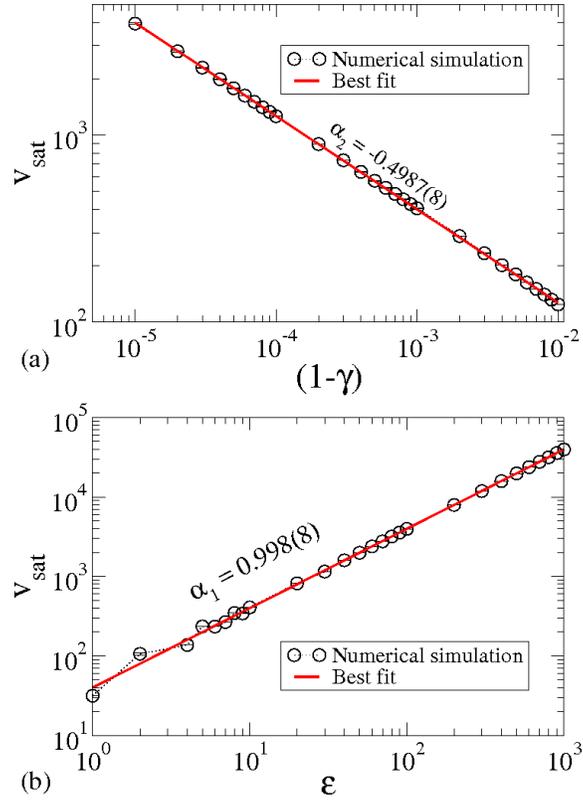}}
\caption{{\it (a) Plot of $v_{\rm sat}$ as a function of $(1-\gamma)$. A
power law fitting gives $\alpha_2=-0.4987(8)\cong-1/2$. (b) Plot of
$v_{\rm sat}$ as a function of $\varepsilon$. A power law fitting furnishes
$\alpha_1=0.998(8)\cong 1$.}}
\label{Fig7}
\end{figure}

The theoretical values for the critical exponents $\alpha_1$ and $\alpha_2$
are given by Eq. (\ref{Eq12}). They are indeed well confirmed from numerical
simulations, as shown in Fig. \ref{Fig7}.

\begin{figure}[htb]
\centerline{\includegraphics[width=0.55\linewidth]{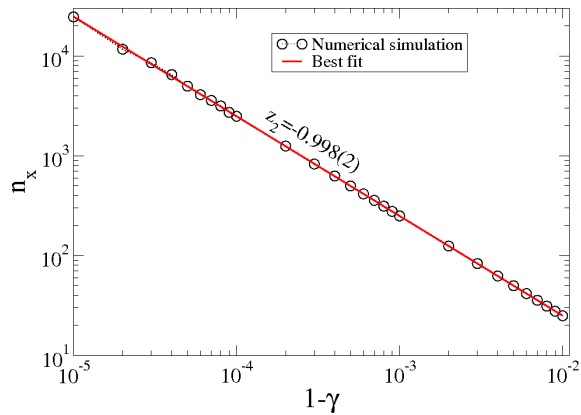}}
\caption{{\it Plot of the crossover $n_x$ as a function of $(1-\gamma)$. A
power law fitting furnishes $z_2=-0.998(2)$ in well agreement with the
result of Eq. (\ref{Eq36}).}}
\label{Fig8}
\end{figure}

Since the exponents $\beta$, $\alpha_1$ and $\alpha_2$ are known, the two
scaling laws given by Eqs. (\ref{Eq34}) and (\ref{Eq36}) can be evaluated to
obtain $z_1$ and $z_2$. When we substitute the numerical values for
$\beta=1/2$, $\alpha_1=1$ and $\alpha_2=-2$ we obtain that $z_1=0$ and
$z_2=-1$. The result $z_1=0$ is not a surprise. Indeed if we look at Fig.
\ref{Fig6}(a) we see that the crossover $n_x$ is the same and is independent
on $\varepsilon$, but do indeed depend on $\gamma$. The
exponent $z_2=-1$ can also be confirmed by numerical simulation. Figure
\ref{Fig8} shows a plot of the crossover $n_x$ as a function of $(1-\gamma)$.
The slope obtained is $z_2=-0.998(2)\cong -1$ in a remarkably well agreement
with the theoretical prediction.

The exponents $z_1$ and $z_2$ can also be obtained from theoretical analysis.
Indeed if we equalling the equation describing the growth of the velocity,
Eq. (\ref{Eq8}) with the equation given the stationary state, Eq.
(\ref{Eq12}), both in dimensionless variables, the isolated $n$ gives the
crossover as function of either $\varepsilon$ and $\gamma$. Then we obtain
\begin{equation}
\sqrt{v^2_0+2n\varepsilon^2}={\sqrt{{(1+\gamma)}\over{{2}}}}{{\varepsilon
} \over {
(1-\gamma)^{{1}\over{2}}}}~,
\label{Eq37}
\end{equation}
and when considering the initial velocity sufficiently small, i.e. $v_0\cong
0$, we end up with
\begin{equation}
n_x={{(1+\gamma)}\over{4}}(1-\gamma)^{-1}~.
\label{Eq38}
\end{equation}
The result given by Eq. (\ref{Eq38}) confirms two things: (i) that the
exponent $z_2=-1$, as obtained by the scaling law (\ref{Eq36}) and confirmed
by numerical simulation from Fig. \ref{Fig8} and; (ii) that the crossover is
independent on $\varepsilon$, leading to $z_1=0$ and as mentioned before.

\begin{figure}[htb]
\centerline{\includegraphics[width=0.55\linewidth]{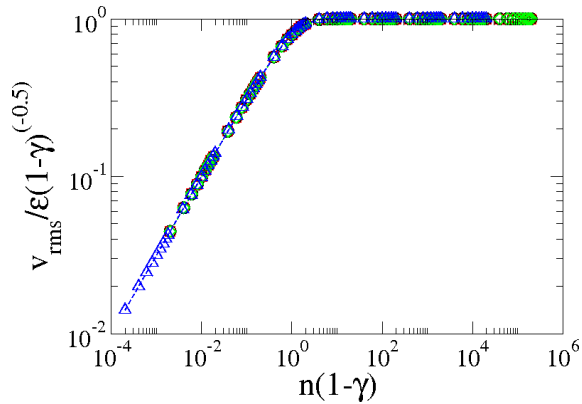}}
\caption{{\it Overlap of all curves shown in Fig. \ref{Fig6}(a) onto a single
and universal plot, after the application of the transformations given by
Eqs. (\ref{Eq39}) and (\ref{Eq40}).}}
\label{Fig9}
\end{figure}

Other property emerging from the critical exponents is the scaling invariance
of the curves shown in Fig. \ref{Fig6}. If the axis of Fig. \ref{Fig6}(a) are
rescaled according to these two transformations namely
\begin{eqnarray}
v&\rightarrow&{{v}\over{\varepsilon^{\alpha_1}(1-\gamma)^{\alpha_2}}}~,
\label{Eq39}\\
n&\rightarrow&{{n}\over{(1-\gamma)^{z_2}}}~,
\label{Eq40}
\end{eqnarray}
all the curves are merged onto a single and universal plot, as shown in Fig.
\ref{Fig9}.

\section{Relation between number of collisions and time}
\label{sec4}

~~~In this section we discuss a way on how to relate the number of collisions
and time. Indeed, from an experimental point of view, measuring the number of
collisions of particles almost massless would be a very difficult task.
Instead of measuring the number of collisions, the easiest parameter is the
time. The main goal of this section is then to relate $n$ and time $t$.

Starting with an experiment, the time can be obtained from
\begin{equation}
t=\Delta t_{c1}+\Delta t_{c2}+\Delta t_{c3}+\ldots +\Delta t_{cn}~,
\label{Eq41}
\end{equation}
where $\Delta t_{ci}$ with $i=1,2,3,\ldots ,n$ gives the interval of
time between two collisions of the same particle. It then gives us an
expression of the type
\begin{eqnarray}
t&=&{{2V_0}\over{g}}+{{2V_1}\over{g}}+{{2V_2}\over{g}}+\ldots
+{{2V_n}\over{g}}~,\nonumber\\
&=&{{2}\over{g}}\sum_{i=0}^{n-1}~V_i.
\label{Eq42}
\end{eqnarray}
If the number of collisions is relatively large, which in the majority of
the situations is, we can approximate the sum of Eq. (\ref{Eq42}) by an
integral of the type
\begin{equation}
t={{2}\over{g}}\int_0^nV(\tilde{n})d\tilde{n}~.
\label{Eq43}
\end{equation}

\begin{figure}[htb]
\centerline{\includegraphics[width=0.55\linewidth]{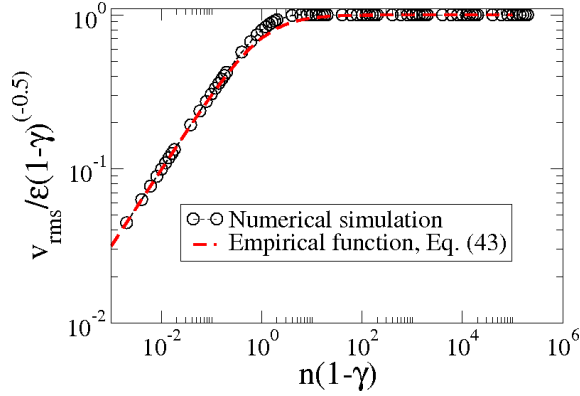}}
\caption{{\it Overlap of a rescaled curve with the empirical function given
by Eq. (\ref{Eq44}).}}
\label{Fig10}
\end{figure}

To make the integral we have to relate indeed $V$ directly with the number of
collisions $n$. This seems to be a difficult task at first sigh. However, the
behaviour shown in Fig. \ref{Fig9} can be described, and it indeed was with
success \cite{Ref35}, by an empirical function of the type
\begin{equation}
f(x)=\left[{{x}\over{1+x}}\right]^{\beta}~,
\label{Eq44}
\end{equation}
where $f$ stands for the vertical axis, $x$ stands for the horizontal axis
and $\beta$ is the accelerating exponent, which for our case here is
$\beta=1/2$. Figure \ref{Fig10}
shows the two curves of rescaled velocity and empirical function given by Eq.
(\ref{Eq44}). We see an astonishing agreement between the two. This is the
link we need to relate velocity $v$ and number of collisions $n$.

With the empirical function at hands, we obtain that
\begin{equation}
{{v}\over{\varepsilon^{\alpha_1}(1-\gamma)^{\alpha_2}}}=\left[{{\left({{n}
\over {(1-\gamma)^{z_2}}}
\right)}\over{\left(1+{{n}\over{(1-\gamma)^{z_2}}}\right)} }
\right]^{\beta
}~.
\label{Eq45}
\end{equation}
Considering the numerical values obtained for the critical exponents, namely
$\alpha_1=1$, $\alpha_2=-1/2$, $z_2=-1$ e $\beta=1/2$, we can rewrite
Eq. (\ref{Eq45}) as
\begin{equation}
{{v}\over{\varepsilon}}\sqrt{1-\gamma}=\sqrt{{{n(1-\gamma)}\over{
1+n(1-\gamma)}}}~.
\label{Eq46}
\end{equation}
Since the dimensionless variables were obtained from $v=\omega
V/g$ and $\varepsilon=\epsilon\omega^2/g$, we can then return to the original
variable $V$ and obtain
\begin{equation}
V(n)={{\epsilon\omega}\over{\sqrt{1-\gamma}}}\sqrt{{n(1-\gamma)}\over{
1+n(1-\gamma)}}~.
\label{Eq47}
\end{equation}

The expression given by Eq. (\ref{Eq47}) can now be used to make the
integration in Eq. (\ref{Eq43}). Then we have
\begin{equation}
t={{2}\over{g}}\int_0^n{{\epsilon\omega}\over{\sqrt{1-\gamma}}}\sqrt{{
\tilde{n}(1-\gamma)}\over{1+\tilde{n}(1-\gamma)}}d\tilde{n}~.
\label{Eq48}
\end{equation}
Using $n^{\prime}=\tilde{n}(1-\gamma)$ and grouping the terms properly we end
up with
\begin{equation}
t={{2\epsilon\omega}\over{g\sqrt{1-\gamma}(1-\gamma)}}\int_0^{n(1-\gamma)}
\sqrt{{n^{\prime}}\over{1+n^{\prime}}}dn^{\prime}~.
\label{Eq49}
\end{equation}

After doing the integral we obtain
\begin{eqnarray}
t&=&{{2\epsilon\omega}\over{g\sqrt{1-\gamma}(1-\gamma)}}\left[-{{1}\over{2}}
\ln(n(1-\gamma))+\sqrt{n^2(1-\gamma)^2+n(1-\gamma)}\right]-
\nonumber\\
&-&{{2\epsilon\omega}\over{g\sqrt{1-\gamma}(1-\gamma)}}\ln\left(1+\sqrt { 1+ {
{ 1 } \over{n(1-\gamma)}} } \right)~.
\label{Eq50}
\end{eqnarray}

For $n\gg 1$, there is only one dominant term in Eq. (\ref{Eq50}), that leads
to
\begin{eqnarray}
t&\cong&{{2\epsilon\omega}\over{g\sqrt{1-\gamma}(1-\gamma)}}[n(1-\gamma)]~,
\nonumber\\
&=&{{2\epsilon\omega}\over{g\sqrt{1-\gamma}}}n~.
\label{Eq51}
\end{eqnarray}
Isolating $n$ we have
\begin{equation}
n={{g\sqrt{1-\gamma}}\over{2\epsilon\omega}}t~.
\label{Eq52}
\end{equation}
Therefore this is the expression we need to obtain the observables as
function of time instead of the number of collisions.

\section{Connections with the thermodynamics}
\label{sec5}

~~~Let us discuss in this section the possible connections with the
thermodynamics for the one-dimensional bouncer model. The first step is the
discussion of the temperature. The connection of the squared velocity with the
temperature comes from the energy equipartition theorem \cite{Ref36,Ref37}.
Indeed, the theorem says that each quadratic term in the expression of the
Hamiltonian (energy) of the system contributes with $K_BT/2$ in the thermal
energy. Here $K_B$ stands for the Boltzmann constant. To apply the theorem,
the system must be in equilibrium, i.e., reached the steady state. However, we
want to describe the evolution of the temperature as function of both $n$ and
time $t$, which may not be at equilibrium yet. We notice however there are
basically two time scales involved in this problem. One of them is the time
the particle spends travelling around (up and down) in between the collisions.
This is a measurable time in the laboratory. The other one, which is several
order of magnitude different, smaller indeed, is the period of an atomic
oscillation, which is governed by the frequency of oscillation $\omega$. Then
between one collision and the other, the atom has completed several thousands
of billions of oscillations. Therefore between collisions, it can be
considered that, from the scale of atomic oscillation, the velocity of the
particle is almost constant, allowing to apply the theorem. In this way we
have
\begin{equation}
{{m\overline{V^2}}\over{2}}={{K_BT}\over{2}}~,
\label{Eq53}
\end{equation}
yielding to
\begin{equation}
\overline{V^2}={{K_BT}\over{m}}~.
\label{Eq54}
\end{equation}

The four important expressions we have then in connection with the
thermodynamics are the squared velocity and the temperature, both written as
function of $n$ and time $t$, as given below. The first one is the squared
velocity as function of $n$
\begin{equation}
\overline{V^2}(n)=V_0^2e^{(\gamma^2-1)n}+{{(1+\gamma)\epsilon^2\omega^2}\over{
2(1-\gamma)}}\left[1-e^{(\gamma^2-1)n}\right]~.
\label{Eq55}
\end{equation}
The second is the temperature as function of $n$ too,
\begin{equation}
T(n)=T_0e^{(\gamma^2-1)n}+{{(1+\gamma)}\over{2}}{{m}\over{K_B}}{{
\epsilon^2\omega^2} \over { (1-\gamma) } }[1-e^{(\gamma^2-1)n}]~.
\label{Eq56}
\end{equation}
The third and fourth are obtained from the Eqs. (\ref{Eq55}) and (\ref{Eq56})
but using the result given by Eq. (\ref{Eq52}). The squared velocity and
temperature as a function of time are then written as
\begin{eqnarray}
\overline{V^2}(t)&=&V_0^2e^{(\gamma^2-1){{g\sqrt{1-\gamma}}\over{
2\epsilon\omega}}t
}+{{(1+\gamma)\epsilon^2\omega^2}\over{2(1-\gamma)}}\left[1-e^{(\gamma^2-1){{
g\sqrt{1-\gamma}}\over{2\epsilon\omega}}t}\right]~,\label{Eq57}\\
T(t)&=&T_0e^{(\gamma^2-1){{g\sqrt{1-\gamma}}\over{2\epsilon\omega}}t}+{{
(1+\gamma)}\over{2}}{{\epsilon^2\omega^2}\over{(1-\gamma)}}{{m}\over{K_B}}
\left [ 1-e^{(\gamma^2-1){{g\sqrt{1-\gamma}}\over{2\epsilon\omega}}t}\right]~.
\label{Eq58}
\end{eqnarray}

Other observable that can be obtained from the system is the entropy. Since
we are considering the measure of the velocity of the particle at the instant
of the collisions, the kinetic energy can then be written as
\begin{equation}
U(t)={{m}\over{2}}\overline{V^2}(t)={{K_B}\over{2}}T(t)~.
\label{Eq59}
\end{equation}

It is known \cite{Ref36} however that
\begin{equation}
{{\partial S}\over{\partial U}}={{1}\over{T}}~,
\label{Eq60}
\end{equation}
that yields to
\begin{equation}
dS={{1}\over{T}}dU~,
\label{Eq61}
\end{equation}
where
\begin{equation}
T(t)={{2}\over{K_B}}U(t)~.
\label{Eq62}
\end{equation}
Then we obtain that
\begin{equation}
\int_{S_0}^SdS^{\prime}={{K_B}\over{2}}\int_{U_0}^U{{dU^{\prime}}\over{U^{
\prime}}}~.
\label{Eq63}
\end{equation}
Doing the integral we have
\begin{equation}
S(t)=\tilde{S}+{{K_B}\over{2}}\ln(U(t))~,
\label{Eq64}
\end{equation}
where $\tilde{S}$ is given by $\tilde{S}=S_0-K_B/2\ln(U_0)$. Equation
(\ref{Eq64}) can also be written in terms of the temperature, i.e.
\begin{equation}
S(t)=\tilde{S}+{{K_B}\over{2}}\ln\left[{{K_B}\over{2}}T(t)\right]~.
\label{Eq65}
\end{equation}

The result given by Eq. (\ref{Eq64}) shows the entropy is dependent on the
energy and grows with the increase of the energy, which is already expected
for an ideal gas \cite{Ref36,Ref37}. Either Eqs. (\ref{Eq64}) and (\ref{Eq65})
must be applied with some care, particularly in the domain of low energy,
implying in low temperature. This is because at such regime, the entropy may
assume negative values. Worst than that is that it can accelerate to $-\infty$
in the limit of $T\rightarrow 0$. Of course in such a limit our approach is
not valid anymore and quantum description must be made.

\section{Discussions}
\label{sec6}

~~~In this section we discuss our results obtained for the steady state
regime. We see from Eq. (\ref{Eq58}) for long enough time, i.e.,
$t\rightarrow\infty$, we obtain
\begin{equation}
T(t\rightarrow\infty)={{(1+\gamma)}\over{2}}{{\epsilon^2\omega^2}\over{
(1-\gamma)}}{{m}\over{K_B}}~.
\label{Eq66}
\end{equation}
The knowledge of the quantities $\epsilon$, $\omega$, $m$ and $T$ allow us to
make an estimation of the restitution coefficient $\gamma$. Therefore
isolating $\gamma$ from Eq. (\ref{Eq66}) we end up with
\begin{equation}
\gamma={{T(\infty)-{{\epsilon^2\omega^2m}\over{2K_B}}}\over{T(\infty)+{{
\epsilon^2\omega^2m } \over{2K_B}}}}~.
\label{Eq67}
\end{equation}

To make an estimation of $\gamma$ we shall consider a light molecule,
Hydrogen ($H_2$) indeed whose atomic mass is $m=2u_a$ where
$u_a=1.660538921\times 10^{-27}kg$. We suppose also the molecule is colliding
with a copper wall at room temperature $T=300K$. From Ref. \cite{Ref32a},
the frequency of oscillation is $\omega=4.49\times 10^{13}Hz$. The amplitude
of oscillation comes from Ref. \cite{Ref32}, but we must give a short
discussion first. 

According to the theoretical models of lattices, a cubic lattice is formed by
atoms placed on each corner of a cube and also with an atom at each face of
the cube. Then the lattice spacing describes the distance between two adjacent
corners of the cube. By knowing the lattice spacing we make the following
assumption. An atom can not vibrate freely. Due to the bonds with other
atoms, it must vibrate only around its equilibrium position. We assume for
doing the estimation of $\gamma$ that each atom oscillates with at most 5
percent of the lattice spacing (which may still be much). Then according to
\cite{Ref32}, $d_l=1.41\times 10^{-10}m$, yielding in $\epsilon=5d_l/100$.

With the numerical values of the parameters at hand, we evaluate Eq.
(\ref{Eq67}) and found a naive estimation for the restitution coefficient of
$\gamma\cong 0.92$.

Using this result to be applied in Eq. (\ref{Eq57}) yields a velocity of
$V_{\rm rms}\cong 310m/s$. The model presented in this paper then succeeded
well to argue on the suppression of Fermi acceleration and making a
temperature finite in a gas. It may also present some applicability to
estimate the restitution coefficient, what should be tested experimentally.
However, it fails to estimate with good accuracy the velocity of a Hydrogen
molecule in a gas at room temperature. Our result furnishes $V_{\rm rms}\cong
V_B/3$ of the velocity measured from a Hydrogen as an ideal gas at room
temperature in 3-D space \cite{Ref38}. The sub index $B$ stands for Boltzmann
velocity.

\section{Conclusions}
\label{sec7}

~~~We have considered in this paper the dynamics of a one-dimensional bouncer
model as an approximation to describe a gas of classical particles moving
along an axis under the influence of gravity. As it is known in the
literature \cite{Ref29}, we confirmed that when a conservative case is
considered, depending on the set of control parameters \cite{Ref2}, unlimited
diffusion is observed for the velocity of the particle, leading to a
phenomenon of Fermi acceleration \cite{Ref31}. However, this unlimited
diffusion is not what one may observe in a laboratory. Therefore, when
inelastic collisions are taken into account, the unlimited diffusion was
suppressed \cite{Ref34} and we have shown two different scenarios evolving
towards the stationary state. For large initial velocity,
we prove that decay of the particle is described by an exponential function
with the speed of the decay depending on the amount of the dissipation,
namely $|(1-\gamma)|$. For very low initial velocity, we have shown the
average velocity grows until bends towards a regime of convergence,
indicating the steady state was reached. The stationary state depends on the
control parameters as $\epsilon(1-\gamma)^{-1/2}$. The crossover marking the
change from growth to the saturation depends on the dissipation
$(1-\gamma)^{-1}$ and is not dependent on $\epsilon$. A scaling law
describing such behaviour was obtained.

The connection with the thermodynamics was made by using the equipartition
theorem \cite{Ref36}. We found expressions for the temperature as well as
squared velocity in variables $n$, denoting the number of collisions of the
particles with the wall, and $t$, which is the time. The later was only
possible to obtain because the scaled velocity was described by an empirical
function \cite{Ref35} relating explicitly velocity $V$ and $n$, as requested
by the procedure. The expression of $T(t)$ so far is more convenient from
experimental point of view and can be measured directly in the laboratory. An
expression for the entropy was also obtained. It has validity only for the
regime of temperature sufficiently large in the sense quantum effects are
negligible.

Finally a comparison of the results obtained from the model was made with
data obtained from \cite{Ref32} and \cite{Ref32a} considering standard
vibrations of atoms as well as frequencies and amplitudes. Our result indicate
a restitution coefficient for the Hydrogen of $\gamma\cong 0.92$ experiencing
collisions with a cooper platform at room temperature. Therefore an
experiment must be made as an attempt to confirm possibly the result
obtained for $\gamma$. As discussed above, our model however fails to predict
with accuracy the average velocity of a gas at room temperature since our
result for 1-D model gives $V_{\rm rms}\cong V_B/3$, for the case considered
in the discussion. Here $V_B$ is the Boltzmann velocity for a 3-D ideal gas.

\section{Acknowledgements}
~~~EDL acknowledges support from CNPq, FUNDUNESP and FAPESP (2012/23688-5),
Brazilian agencies. This research was supported by resources supplied by the
Center for Scientific Computing (NCC/GridUNESP) of the S\~ao Paulo State
University (UNESP). ALPL thanks CNPq for financial support and is also
grateful for the kind hospitality of School of Mathematics of University of
Bristol - UK for the time spent as PhD Sandwich supported by CAPES CsF -
0287-13-0 (Brazilian agency), which is also acknowledged.

\section*{Appendix 1 - Standard map}

~~~The standard mapping is written as
\begin{equation}
\left\{\begin{array}{ll}
I_{n+1}=I_n+K\sin(\theta_n)\\
\theta_{n+1}=[\theta_n+I_{n+1}]~{\rm mod~(2 \pi)}
\end{array}
\right.,
\label{Eq_Ap1}
\end{equation}
where $K$ is a control parameter that controls two different types of
transition: (1) Integrability with $K=0$ to non-integrability for $K\ne 0$
and; (2) Transition from local for $K<K_c=0.9716\ldots$ to globally chaotic
behaviour $K\ge K_c$. A global chaotic behaviour means the invariant spanning
curves separating the phase space in different portions are all destroyed
letting the chaotic sea to diffuse unbounded.

\end{document}